\documentstyle[psfig]{mn0}

\def\lta{\mathrel{\rlap{\raise 0.511ex \hbox{$<$}}{\lower 0.511ex \hbox{$\sim$}}}}
\def\gta{\mathrel{\rlap{\raise 0.511ex \hbox{$>$}}{\lower 0.511ex \hbox{$\sim$}}}}

\begin{document}

\title[GRB 080319B]{What did we learn from gamma-ray burst 080319B ?}

\author[Kumar \& Panaitescu]{P. Kumar$^{1}$ and A. Panaitescu$^{2}$ \\
    $^{1}$ Department of Astronomy, University of Texas, Austin, TX 78712, USA  \\
    $^{2}$ Space Science and Applications, Los Alamos National Laboratory, 
          Los Alamos, NM 87545, USA }

\maketitle

\begin{abstract}
\begin{small}
 The optical and gamma-ray observations of GRB 080319B allow us to provide 
a broad-brush picture for this remarkable burst. The data indicate that the 
prompt optical and gamma-ray photons were possibly produced at the same location 
but by different radiation processes: synchrotron and synchrotron self-Compton, 
respectively (but we note that this interpretation of the gamma-ray data faces 
some difficulties). We find that the burst prompt optical emission was produced 
at a distance of $10^{16.3}$ cm by an ultra-relativistic source moving at Lorentz 
factor of $\sim 500$. A straightforward inference is that about 10 times more 
energy must have been radiated at tens of GeV than that released at 1 MeV. 
Assuming that the GRB outflow was baryonic and that the gamma-ray source was 
shock-heated plasma, the collimation-corrected kinetic energy of the jet powering 
GRB 080319B was larger than $10^{52.3}$ erg. The decay of the early afterglow 
optical emission (up to 1 ks) is too fast to be attributed to the reverse shock 
crossing the GRB ejecta but is consistent with the expectations for the 
``large-angle" emission released during the burst. The pure power-law decay of 
the optical afterglow flux from 1 ks to 10 day is most naturally identified with 
the (synchrotron) emission from the shock propagating into a wind-like medium. 
However, the X-ray afterglow requires a departure from the standard blast-wave 
model. 
\end{small}
\end{abstract}


\section{Introduction}

 Our understanding of gamma-ray bursts (GRBs) has improved tremendously in the 
last ten years. Observations have firmly established that GRBs and their afterglows 
arise from highly relativistic (Taylor et al 2004), collimated outflows (Frail 
et al 2001, Panaitescu \& Kumar 2002). At least a few long-duration GRBs are 
associated with the collapse of massive stars (Galama et al 1998, Della Valle 
et al 2003, Hjorth et al 2003, Malesani et al 2004), but less than a few percent 
of supernovae of Type Ib/c give rise to GRBs (Soderberg et al 2004). A fraction 
of short-duration GRBs are associated with old stellar populations (Nakar 2007).

 Among the fundamental questions that remain unanswered is the generation of
$\gamma$-rays and the composition of the relativistic jet (baryonic, e$^\pm$ and/or 
magnetic). The simultaneous monitoring of the optical and $\gamma$-ray emissions 
of GRB 080319B can provide insight into the mechanism for $\gamma$-ray generation, 
as described in the next section.

 Studies of the multiwavelength prompt emission are still incipient because of
the rarity of a good temporal coverage of the optical counterpart. So far, the 
prompt optical emission was observed 
$(i)$ to be decoupled from that at $\gamma$-rays (GRB 990123 -- Akerlof et al 1999), 
$(ii)$ to contain a fluctuating component that tracks the burst and a smooth one 
    peaking after the burst (GRB 050820A -- Vestrand et al 2006), or
$(iii)$ to fluctuate in phase with the burst (GRB 041219A -- Vestrand et al 2005).
GRB 080319B is of the last type.

\section{Prompt radiation}

\subsection{Synchrotron self-Compton model}

 GRB 080319B was the brightest burst detected by Swift. It was also seen by 
the Konus satellite, which measured its fluence in the 20 keV -- 7 MeV band, 
$\Phi_\gamma = (5.7 \pm 0.1) \times 10^{-4}$ erg cm$^{-2}$, corresponding to
and isotropic energy release of $E_\gamma = 1.3 \times 10^{54}$ erg (Golenetskii 
et al 2008), for the burst redshift $z=0.937$ (Vreeswijk et al 2008, Cucchiara \&
Fox 2008). The time-averaged spectrum during the burst peaked at 650 keV. 
The Konus spectrum was $F_\nu \propto \nu^{0.18 \pm 0.01}$ below its peak at 
650 keV and $F_\nu \propto \nu^{-2.87 \pm 0.44}$ above it.

\begin{figure}
\psfig{figure=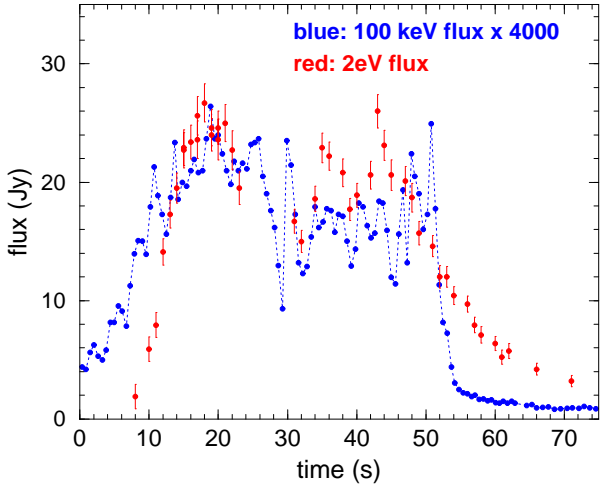,width=8cm}
\caption{ Prompt gamma-ray and optical light-curves of GRB 080319B. 
  The high energy flux was multiplied by 4000 to allow comparison with the much 
  larger optical flux. The burst gamma-ray emission was monitored by the Swift 
  BAT instrument at 15--150 keV (data shown here are courtesy of T. Sakamoto 
  and S. Barthelmy). The optical counterpart emission was measured by the 
  Tortora collaboration (Karpov et al 2008). The finer time-sampling of the 
  burst light-curve reveals more variability than can be seen in the optical 
  emission, still the two light-curves appear correlated: the three optical 
  peaks at 18, 35, and 43 s correspond to peaks of the GRB emission. }
\label{prompt}
\end{figure}

 The peak optical flux during the burst was $V=5.4$ mag (Karpov et al 2008), 
detectable with unaided eye for about 30 s under appropriate conditions. 
The optical and the gamma-ray light-curves of GRB 080319B shown in Figure 
\ref{prompt} are mildly correlated, suggesting that both radiations originated 
from the same source. The optical-gamma-ray correlation function for the entire 
prompt emission has a maximum at lags between 0s and 5s, that being mostly 
the result of the contemporaneous rise at 0-18 s and fall at 43-60 s. 
For the middle part of prompt emission, the optical-gamma-ray correlation 
function still has a peak at lags between -1s to 3s, but is less prominent.

 The time-lag between the (GRB) emission produced in (say) internal shocks and the 
(optical) emission released by the blast-wave caused by the ejecta interaction
with the ambient medium can be arbitrarily small, thus the optical and $\gamma$-ray 
fluctuations could be correlated (as seen for GRB 080319B) even though they arise 
from different sources. However, the short duration $\delta t$ of the optical pulses 
(with $0.2 \lta \delta t / t \lta 0.5$) and the lack of an increasing optical pulse 
duration throughout the burst (instead, a decrease is apparent), as expected for an
increasing source radius, argue against an origin of the optical counterpart in the 
external shock (either the reverse-shock crossing the ejecta or the forward-shock 
propagating in the circumburst medium). This makes it more likely that the two prompt
emissions are, in fact, arising from same source.

 The extrapolation of the gamma-ray spectrum (which peaks at 650 keV and has
a peak flux of about 7 mJy) to the optical band under-predicts the 
observed optical counterpart flux (which is around 20 Jy) by almost 4 orders 
of magnitude. This indicates that, although from same source, the optical and 
gamma-ray radiation processes are different. A natural possibility is 
that optical photons were generated via the synchrotron emission and gamma-rays 
via inverse-Compton scatterings of that synchrotron emission (i.e. {\sl synchrotron 
self-Compton} mechanism). 

 The peak energy of the synchrotron spectrum ($\nu_i$) should be below optical, to 
account for the optical spectrum of $F_\nu \propto \nu^{-2/3}$ (Wo\'zniak et al 2008):
measured by Raptor after 80 s. We parameterize the synchrotron peak photon energy 
as $\nu_{io}$eV ($\nu_{io} \lta 1$), above which the optical counterpart spectrum 
has a spectral index $\beta_o$ (i.e. $F_\nu \propto \nu^{-\beta_o}$). 
The measured peak energy of the $\gamma$-ray burst spectrum, which is the 
inverse-Compton scattering of synchrotron photons, implies that the radiating 
electrons have a random Lorentz factor $\gamma_e = (650\, {\rm keV}/5\, \nu_{io}\, 
{\rm eV})^{1/2} \sim 360\, \nu_{io}^{-1/2}$, where the factor 5 accounts for the 
scattering by a population of electrons with a power-law distribution with energy 
above $\gamma_e$. The second inverse-Compton scattering component should peak at 
$5\gamma_e^2 * 650\, {\rm keV} \sim 400\, \nu_{io}^{-1}$ GeV and its fluence should 
be larger than that of the gamma-rays by a factor equal to the gamma-ray to optical 
fluence ratio: $Y_1 = (650\, {\rm keV} * 7 {\rm mJy})/ (2\,{\rm eV} * 20\, {\rm mJy} 
* \nu_{io}^{1-\beta_o}) \sim 100$ (this ratio is the "Compton parameter").
Hence, the fluence of the $\sim 400$ GeV emission accompanying GRB 080319B should 
have been $\Phi_{GeV} = Y_1 \Phi_\gamma \sim 0.05$ erg cm$^{-2}$; the GLAST/LAT 
instrument would collect thousands photons at 0.1--100 GeV during a burst as bright 
as GRB 080319B.

 For the isotropic-equivalent gamma-ray output of GRB 080319B of $E_\gamma \sim 
10^{54}$ erg, the Compton parameter above implies a GeV output of $E_{GeV} \simeq 
Y_1 E_\gamma \sim 10^{56}$ erg. The energy release is smaller if the second 
inverse-Compton scattering is in the Klein-Nishina regime. The total energy 
release is reduced further by a few orders of magnitude if the fireball is tightly 
collimated (i.e. the explosion energy is released in a narrow jet). These reduction 
factors are estimated below.

 As the peak of the synchrotron spectrum is at $\nu_i = 10^{6.6} \, (z+1)^{-1}
B\Gamma \gamma_e$ Hz, where $B$ is the source's magnetic field and $\Gamma$ its bulk 
Lorentz factor, the inferred random Lorentz factor $\gamma_e$ of the radiating 
electrons and the peak frequency of the synchrotron spectrum imply that $B \Gamma 
\sim 10^3 \nu_{io}^2$ Gauss. For a burst redshift $z=0.94$, the peak flux of the 
synchrotron emission is $F_p = 10^{-56.2} \Gamma B N_e$ Jy, where $N_e$ is 
the number of radiating electrons, thus observed synchrotron optical peak flux of 
$\sim 20$ Jy requires that $N_e \sim 10^{54.5}\, \nu_{io}^{-\beta_o-2}$. 
The optical thickness of the source to Thomson scattering ($\tau_e$) is the ratio 
of the inverse-Compton peak flux (at 650 keV) 
and the synchrotron peak flux: $\tau_e \sim 10^{-3.5}\, \nu_{io}^{\beta_o}$. 
Using that $\tau_e = \sigma_{Th} N_e/(4\pi R_\gamma^2)$, where $\sigma_{Th}$ is the 
electron cross-section for Thomson scattering, it follows that the GRB source radius 
is $R_\gamma \sim 10^{16.3}\, \nu_{io}^{-\beta_o-1}$ cm. This distance is a 
factor $\sim 100$ larger than that expected (eg. Piran 2005) for internal shocks. 

\begin{figure}
\psfig{figure=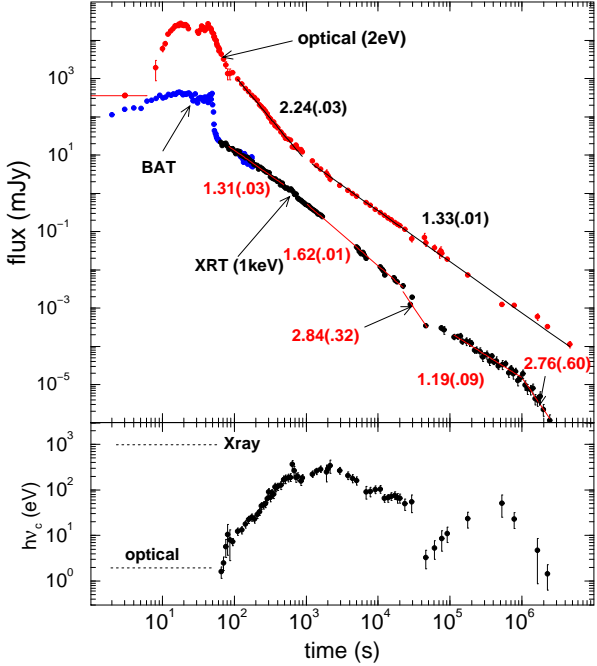,width=8cm}
\caption{ {\sl Top panel}: prompt and afterglow X-ray and optical light-curves 
  of GRB 080319B. The BAT measurements of the prompt 15--150 keV emission have
  been shifted vertically to match the XRT flux at 65--180 s. 
  The XRT 0.3--10 keV afterglow measurements are from the Evans et al (2007) 
  site and the 1 keV flux was calculated using the spectrum $F_\nu \propto 
  \nu^{-0.92 \pm 0.07}$ reported by Racusin et al (2008).
  Optical measurements are from Bloom et al (2008) and have been supplemented
  with the measurements after 1 day reported in GCN 7535 (D. Perley et al),
  7569 (N. Tanvir et al), 7621 (N. Tanvir et al), and 7710 (A. Levan et al).
  Power-law fits to segments of each light-curve are shown. Numbers indicate the 
  index (and uncertainty) of the power-law decay exponent $\alpha$ defined by 
  $F_{o,x} \propto t^{-\alpha}$. 
  {\sl Bottom panel}: evolution of the cooling break frequency $\nu_c$ required
  to accommodate the afterglow optical and X-ray fluxes with the same spectral
  component. This frequency is calculated from the slope of the X-ray continuum,
  taking into account that below $\nu_c$ the slope of the synchrotron spectrum 
  is smaller by 1/2 than above it. Extinction of the optical emission by dust
  in the host galaxy is negligible (Bloom et al 2008). }
\label{aglow}
\end{figure}

 The very steep decline of gamma-ray and optical flux at the end of the burst 
suggests that the source turned-off quickly. The flux decline cannot be faster
than the limit provided by the photons emitted from source regions outside the 
$\Gamma^{-1}$ opening area moving toward the observer. This ``large-angle" emission 
arising from the fluid moving at an angle larger than $\Gamma^{-1}$ yields a flux 
decay $F_\gamma \propto t^{-\beta_\gamma-2}$ (Fenimore \& Sumner 1997, Kumar \& 
Panaitescu 2000). Since the burst emission has $\beta_\gamma \sim 0$ in the 15--150 
keV BAT window (Cummings et al 2008), the large-angle emission should decay as 
$F_\gamma \propto t^{-2}$, i.e. much slower than the fall-off displayed by the burst 
emission at 50--55 s (Figure \ref{aglow}). 
The observed decay of the burst tail can be reconciled with the upper limit set by 
the large-angle emission if the last GRB pulse is timed not since the burst trigger 
(e.g. Fan \& Wei 2005) but since a reference time $t_0 = 48$ after it, when the 
relativistic ejecta producing the last GRB pulse must have been released.

 This means that the timescale of the GRB pulse decay is $\delta t \sim 3$ s. 
Taking into account that $\delta t = (z+1) R_\gamma /(2c\Gamma^2)$, the inferred 
burst radius implies a source Lorentz factor $\Gamma \sim 500 \, \nu_{io}^ 
{-(\beta_o+1)/2}$. Then, from $B\Gamma \sim 10^3 \nu_{io}^2$ G, we find that the 
magnetic field in the GRB source is $B \sim 2\, \nu_{io}^{(\beta_o+5)/2}$ G. 

 The GRB source radius can be used to constrain the location of the synchrotron 
peak frequency ($\nu_{io}$). The lack of a progressive lengthening of the burst 
pulse throughout the burst indicates that the GRB source is not decelerating. 
The radius at which the ejecta are decelerated by their interaction with the wind 
expelled by the Wolf-Rayet progenitor is $R_d \sim 10^{17}\, \nu_{io}^{\beta_o+1}$ cm, 
provided that the kinetic energy of the ejecta is similar to the radiation output. 
The condition $R_\gamma < R_d$ leads to $\nu_{io} \gta 0.5$. At the same time, 
at the end of the burst, the peak of the synchrotron spectrum cannot be well above 
1 eV (i.e. $\nu_{io} \lta 3$) because that would bring the synchrotron self-absorption 
frequency above the optical and yield a constant optical flux from the large-angle 
emission, which is inconsistent with observations of a decaying optical flux after 
the burst end. Thus, to a good approximation, $\nu_{io} \sim 1$, implying that the
time averaged optical spectrum during the burst should have been rather flat. 

 For the inferred $\gamma_e$ and $\Gamma$, the second inverse-Compton scattering 
occurs just above the Klein-Nishina limit, where the electron scattering cross-section
is $\sim 0.4 \sigma_{Th}$. Therefore the twice-scattered photon takes all the electron
energy, the peak energy of the 2nd inverse-Compton component being at $\Gamma \gamma_e 
m_e c^2 (z+1)^{-1} \sim 50$ GeV and the Compton parameter for the 2nd scattering being 
$Y_2 = 0.4\, \tau_e \gamma_e m_e c^2 /[\,650\, (z+1) \linebreak \Gamma^{-1}\, {\rm keV}] 
\sim 10$ (i.e. 10 times smaller than for the first scattering). 
 Thus, the isotropic-equivalent total radiation output of GRB 080319B was $E_{rad} 
= Y_2 E_\gamma \sim 10^{55}$ erg. 

 From the inferred $B$, $\Gamma$, and $Y$ (for first and second scattering), it can be 
shown that the cooling electron Lorentz factor $\gamma_c = 6 \pi m_e c^2 \Gamma /
(Y_1 Y_2 R_\gamma B^2) \sim 150$, thus $\gamma_c < \gamma_e$ and the GRB electrons 
are fast-cooling (Sari, Piran \& Narayan 1998). This has two consequences. First, 
since the GRB spectrum does not exhibit a $F_\nu \propto \nu^{-1/2}$ spectrum below 
its peak, as expected for radiatively-cooling electrons, a re-acceleration process 
must be active during the burst.
Second, at the end of burst, when electrons are no longer accelerated, the peak
energy of the inverse-Compton spectrum falls fast below the BAT window and the
emission from the $\Gamma^{-1}$ region moving toward the observer decreases on 
a timescale shorter than $\delta t$, allowing the large-angle emission to become 
dominant (i.e. initial assumption that the GRB tail is the large-angle emission is 
self-consistent).

 If the GRB ejecta were baryonic with one proton per electron, then the energy stored 
in protons were larger by that of electrons by a factor of at least $m_p/(\gamma_e m_e) 
\sim 5$, implying that the total GRB outflow energy is at least 5 times larger than 
the radiated energy. Thus the total isotropic-equivalent energy release in this GRB 
explosion must have been $E_k \gta 5 E_{rad} \sim 10^{55.7}$ erg.

\subsection{Limitations/drawbacks of the model}
\label{lim}

 Although the synchrotron self-Compton model is a {\it natural} interpretation 
of the optical and gamma-ray prompt emission from GRB 080319B, it suffers 
from a few, possibly serious, problems. 
 
 One of the problems is the incompatibility between the observed and expected 
gamma-ray spectrum. In the synchrotron-self Compton model, the synchrotron (optical) 
and inverse-Compton (gamma-rays) spectra are closely related. The model parameters 
derived above lead to an upscattered self-absorbed photon energy of $\sim 100$ keV, 
below which the inverse-Compton spectrum should be $F_\nu \propto \nu$ (Panaitescu 
\& M\'esz\'aros 2000). In contrast, Golenetskii et al (2008) measure for GRB 080319B
a softer $F_\nu \propto \nu^{0.2}$ gamma-ray spectrum down to 20 keV. A similar
inconsistency between the self-Compton model and gamma-ray observations is 
encountered for GRB 990123, which we (Panaitescu \& Kumar 2007) suggested that
could be explained if the magnetic field decays (Rossi \& Rees 2003) on a length-scale 
much larger than the plasma skin depth.

 Furthermore, in the large-angle emission interpretation of the early, steep optical 
decay (\S\ref{lae}), the optical spectral slope $\beta_o = 2/3$ measured by Wo\'zniak 
et al (2008) during that decay is just the slope above the peak of the synchrotron
spectrum of the prompt emission. However, that is inconsistent with the spectral
slope of the inverse-Compton emission above the GRB spectral peak, which Golenetskii 
et al (2008) report to be much softer, $\beta_\gamma = 2.9 \pm 0.4$. The possible
solution to this problem is that the early optical afterglow emission of GRB 080319B, 
which we identify with the large-angle emission released during the burst, is dominated 
by that from a single pulse whose spectrum is harder above its spectral peak 
than that of the burst-averaged gamma-ray spectrum. Taking onto account that
the decay of the large-angle emission from a pulse lasting $\delta t$, peaking at $t_p$, 
and of peak flux $F_p$, is $F_{lae} (t) = F_p [(t-t_p)/ \delta t]^{-2-\beta_o}$, it seems
plausible that the large-angle emission is dominated by the pulse with the hardest
spectral slope $\beta_o$. 

 Another potential problem for the synchrotron self-Compton interpretation of GRB
080319B's prompt emission is that the gamma-ray light-curve exhibits a higher variability
than the optical counterpart, as one would expect the inverse-Compton gamma-ray emission
to have less fluctuations than the synchrotron seed photon field.

 We note that the rise of the optical counterpart emission lagging that of the gamma-rays
could be due to the optical band being below the synchrotron self-absorption frequency 
at early times. In this case, the adiabatic cooling of ejecta electrons leads to a steep 
rise ($t^3$), consistent with that displayed by the optical emission at 8--15 s.
At the other end, the optical counterpart emission turned-off before the gamma-rays. 
That the last three GRB pulses are not accompanied by optical emission may be due
to that their synchrotron spectrum peaked sufficiently below the optical.

\section{Afterglow Emission}

\subsection{Steep, early optical decay ($t < 1$ ks)}
\label{lae}

\begin{figure}
\centerline{\psfig{figure=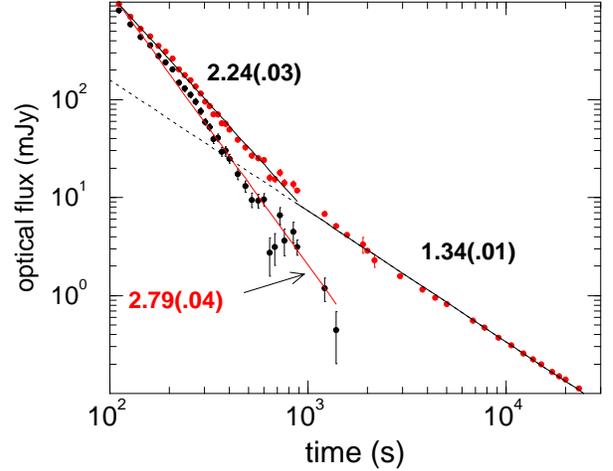,width=8cm}}
\caption{ Subtraction of the extrapolation (solid black line) to earlier times 
  of the afterglow optical emission after 1 ks reveals that the excess of  
  optical flux before 1 ks has a decay $F_o \propto t^{-2.8}$ (red line).  }
\label{early}
\end{figure}

 After the burst, the optical emission has a spectral slope $\beta_o \simeq 0.65$
and displays a steep decay, $F_o \propto t^{-2.8}$ up to 1 ks, as shown in Figure 
\ref{early}, where the contribution of the power-law decaying emission after 1 ks, 
$F_o \propto t^{-1.3}$, was subtracted from the earlier optical flux. 
Assuming that the burst ejecta is in 
pressure equilibrium with the ambient medium shocked by the forward shock, whose 
structure is described by the Blandford-McKee solution, we find that the steepest 
decay for the synchrotron emission from the ejecta electrons, after they have been 
energized by the reverse-shock (i.e. as they cool adiabatically), is obtained for 
a wind-like medium and an ejecta shell thickness increasing as $R/\Gamma_{rs}$, 
with $R$ the shock radius and $\Gamma_{rs}$ its Lorentz factor. That decay, 
$\alpha_{rs} = 1.47\beta_o + 0.80 = 1.75$ (see equation 55 of Panaitescu \& Kumar 
2004) is too slow to account for the decay measured for GRB afterglow 080319B at
100--1000 s (the results of Sari \& Piran 1999 and Kobayashi \& Sari 2000 for the 
reverse-shock emission give a steeper decay, up to $\alpha_{rs} = 2.0$, which is 
still too slow to account for the observed $F_o \propto t^{-2.8}$).  

 Instead, the optical light-curve decline is consistent with the large-angle emission 
model, for which $F_o \propto t^{-2-\beta_o}$ is expected. Therefore, we suggest 
that the optical afterglow emission between 50 s and 1 ks arises from the same source 
that produced $\gamma$-ray photons.

\subsection{Afterglow emission after 1 ks}

 The slower decay of the optical flux observed after 1 ks, $F_o \propto t^{-1.33 \pm 
0.01}$ (Figure \ref{aglow}), is most naturally attributed to the emission from 
the forward-shock driven into the ambient medium (e.g. Paczy\'nski \& Rhoads 1993, 
M\'esz\'aros \& Rees 1997) by the burst ejecta overtaking the large-angle burst emission. 
The optical flux decay ($F_o \propto t^{-\alpha_o}$) and spectral slope $F_\nu \propto 
\nu^{-0.63 \pm 0.07}$ at 90 ks (Bloom et al 2008) satisfy $\alpha_o - 1.5 \beta_o = 
0.39 \pm 0.10$, which is consistent only with a circumburst medium density of $r^{-2}$ 
radial structure and cooling frequency of the forward-shock synchrotron spectrum above 
optical (for which $F_o \propto t^{-(3\beta_o+1)/2}$ is expected -- Chevalier \& Li 
1999). A homogeneous circumburst medium, for which $\alpha_o - 1.5 \beta_o = 0$ or
-0.5 (depending on the location of the cooling break) is inconsistent with the
observations at the 4$\sigma$ level.

 The X-ray light-curve should exhibit the same decay as the optical if the cooling 
break frequency is not between optical and X-ray or, owing to a wind-like medium, a 
slower one with decay index $\alpha_x = \alpha_o - 1/4$ if the cooling frequency is 
between these domains. The former could be the case for the early (before 1 ks) X-ray 
light-curve, whose decay is compatible with that seen in the optical after 1 ks,
but neither expectation is satisfied by the X-ray and optical decays after 1 ks, 
for which $\alpha_x = \alpha_o + 1/3$. The same inconsistency with the standard 
blast-wave model expectations is also illustrated in the lower panel of Figure 
\ref{aglow}, which shows that, if the optical and X-ray afterglow emissions were 
attributed to the same spectral component, then the cooling break would have to 
exhibit an non-monotonic evolution incompatible with the blast-wave model, in general, 
and with the $\nu_c \propto t^{1/2}$ evolution expected for wind medium, in particular. 

 The non-monotonic evolution of the cooling break shown in Figure \ref{aglow}
suggests a departure from the basic assumptions of the standard blast-wave model,
such as a non-monotonically evolving magnetic field energy parameter (on which
the cooling frequency has the strongest dependence). Alternately, the optical and 
X-ray afterglow emissions may have (somewhat) different origins. For instance, 
the optical flux, which exhibits a single power-law decay after 1 ks, could be 
attributed to the synchrotron emission from the forward shock, while the X-ray flux,
which displays 5 power-law segments, could arise from reprocessing of the forward-shock 
emission by scattering off a lagged part of the relativistic outflow (Panaitescu 2008). 

 For GRB ejecta collimated into a jet of half-angle $\theta_j$, the afterglow
light-curve should exhibit a steepening to a decay faster than $t^{-1.5}$ 
when the jet edge becomes visible to the observer. Using the arrival time of 
photons emitted at angle $\Gamma^{-1}$ and for a circumburst medium of density 
typical for Galactic Wolf-Rayet stars, the jet-break time is $t_b = 200\, (z+1) 
\, (E_k/10^{55.7}) (\theta_j/0.1)^4$ d. Since the optical light-curve of GRB 080319B
afterglows does not exhibit a jet-break during the first 10 days (Figure \ref{aglow}), 
the jet opening is $\theta_j \gta 2$ deg. Therefore, the collimation-corrected 
kinetic energy of the GRB outflow is $E_{jet} = E_k (\theta_j^2/4) \gta 10^{52.3}$ erg.

\section{Conclusions}

 The modest correlation between the optical and gamma-ray light-curves of GRB 080319B 
suggests that these emissions may be from the same source. That the prompt optical 
flux lies well above the extrapolation of the burst spectrum to optical indicates
that the two emissions arose from different radiation processes: optical emission 
was generated through synchrotron process and $\gamma$-rays were due to inverse-Compton 
scattering of the synchrotron photons. As shown by (Kumar \& McMahon 2008), a bright 
optical emission accompanying $\gamma$-rays is a generic prediction of the synchrotron 
self-Compton model for GRBs. However, some features of the gamma-ray data for GRB 080319B 
are not accounted for by that model, as discussed in \S\ref{lim}.

 Synchrotron self-Compton emission may have also been observed during GRB 990123 
(Panaitescu \& Kumar 2007). For the bright optical counterpart of GRB 080319B, 
we find that the spectrum of the synchrotron prompt emission must have peaked in 
the optical, while for GRB 990123 that peak was slightly above optical, which can
explain (at least in part) why these optical counterparts were so bright.

 Within the synchrotron self-Compton model for the prompt emission and without
any assumption about the dissipation mechanism, we have shown that the $\gamma$-ray 
and optical measurements of GRB 080319B imply a source radius of $10^{16.3}$ cm, 
moving at Lorentz factor 500. The energy of the 2nd inverse-Compton scattering, which 
should have peaked at $\sim 50$ GeV, must have been 10 times larger than that of the 1st 
inverse-Compton component at sub-MeV energies. With the lower limit on the half-angle
of the GRB outflow set by the lack of an afterglow optical light-curve break until 
10 d, and assuming a baryonic, shock-heated $\gamma$-ray source, we estimate that 
the beaming-corrected kinetic energy of outflow powering GRB 080319B was higher than 
$10^{52.3}$ erg (twice larger for a double-sided jet), which provides an important 
constraint for the collapsar model (MacFadyen \& Woosley 1999, Woosley \& Bloom 2006) 
for long bursts.

 We find that the optical light-curve decay at 
0.1--1 ks is too fast to be attributed to the external reverse-shock propagating 
into the GRB ejecta. Instead, the early optical decay and spectral slope are 
consistent with the large-angle emission released during the burst and arriving
at observer later, due to the spherical curvature of the emitting surface.

 The decay of the late-time ($t \gta 1$ ks) optical afterglow flux is consistent
with an origin in the forward-shock energizing a wind-like circumburst medium,
a homogeneous medium being "ruled out" at 4$\sigma$. However, if the optical and 
X-ray afterglow fluxes arose from same mechanism, an unusual evolution of the 
cooling frequency would be required, which may be indicative of evolving blast-wave 
microphysical parameters. Alternatively, the X-ray afterglow emission may be affected 
by reprocessing of the optical forward-shock photons through (bulk and, possibly,
inverse-Compton) scattering by a relativistic, pair-rich outflow located behind
the forward shock. In this model, the scattering outflow results 
from continued accretion onto the central black-hole, if such an accretion can be 
maintained for source-frame timescales comparable to the observer-frame duration 
of the X-ray afterglow. 

\section*{Acknowledgments}
 The authors are grateful to the referee for his comments.
 AP acknowledges the support of the US Department of Energy through the LANL/LDRD 
20080039DR program and of NASA Swift GI grant NNG06EN001 for this work.

\end{document}